\documentclass[12pt]{article}
\usepackage{amsmath}
\begin{document}
\begin{flushright}
KANAZAWA-11-20\\
January, 2012
\end{flushright}
\vspace*{1cm}

\begin{center}
{\Large\bf  Extension of a radiative neutrino mass model based on a
 cosmological view point}
\vspace*{1cm}

{\Large Daijiro Suematsu}\footnote{e-mail:~suematsu@hep.s.kanazawa-u.ac.jp}
\vspace*{1cm}\\

{\it Institute for Theoretical Physics, Kanazawa University, 
\\ Kanazawa 920-1192, Japan}
\end{center}
\vspace*{1.5cm} 

\noindent
{\Large\bf Abstract}\\
We consider an extension of the radiative neutrino mass model 
at TeV regions so as to give the origin for inflation of the universe. 
This extension also gives a consistent explanation for
both the origin of baryon number asymmetry and dark matter. 
A small scalar coupling which plays a crucial role in the neutrino 
mass generation in the original model may be related to parameters 
which are control inflation.
\newpage
\section{Introduction}
The radiative neutrino mass model proposed in \cite{ma} is a simple 
but interesting possibility which can take the place of the 
canonical seesaw model 
for neutrino masses.  It could give the explanation for not only the 
neutrino mass problem but also the origin of dark matter (DM) 
\cite{raddm1,raddm2,raddm3}.\footnote{The supersymmetric extension of the
model is proposed in \cite{susyrad1,susyrad2}.}
A $Z_2$ symmetry to forbid the tree level neutrino mass
also guarantees the stability of the lightest $Z_2$ odd field.
As a result, the reason of the smallness of neutrino masses and the existence of
DM are closely related in this model.
Although the model has these noticeable features, baryon number asymmetry 
in the universe seems not to be easily explained in this framework 
unfortunately.
The ordinary thermal leptogenesis \cite{fy,leptg} 
seems not to work for the generation of
sufficient lepton number asymmetry in a consistent way with the relic
abundance of DM without some modification \cite{radlept}. 

In this model the lightest right-handed neutrino is a most promising 
DM candidate. In that case both its relic abundance 
and small neutrino masses require $O(1)$ neutrino Yukawa couplings 
in general \cite{raddm1,raddm2,raddm3}. 
This allows the model to cause the large CP asymmetry in the decay of
the right-handed neutrinos even though their masses are of $O(1)$~TeV.
However, the same neutrino Yukawa couplings could cause a large washout of the
generated lepton number asymmetry through the lepton number violating
scattering processes.
As a result, the thermal leptogenesis is not easy to generate sufficient 
lepton number asymmetry in a consistent way with the neutrino
oscillation data and the DM abundance at least in the simplest form 
of the model. Non-thermal leptogenesis might give a consistent scenario
for the origin of the baryon number asymmetry in this model \cite{ad}.

In this paper we propose a simple extension to solve this
fault of the model based on a cosmological viewpoint. 
We extend the model to make it connect with inflation of the
universe and also generate the 
lepton number asymmetry through the inflaton decay.\footnote{The similar
idea has been proposed in the inflation scenario in which the
right-handed sneutrino plays a role of inflaton \cite{sr-inf}.} 
Since the lepton number violating effect could be separated 
from the neutrino mass generation,
both the reduction of DM relic abundance and the washout of lepton number
asymmetry are reconciled for the same neutrino Yukawa couplings.
As a bonus, we have an explanation for the smallness of a scalar 
coupling which is crucial for this radiative neutrino mass generation.  

\section{Extension of the model}
The model considered here is a simple model for the radiative neutrino 
mass generation \cite{ma}.
This model is an extension of the standard model (SM) with three 
right-handed singlet fermion $N_{R_i}$ and an inert doublet scalar $\eta$. 
These new fields are supposed to have odd parity of an assumed 
$Z_2$ symmetry, under which all SM contents have even parity.
Invariant Yukawa couplings and scalar potential related to these new
fields are summarized as
\begin{eqnarray}
-{\cal L}_y&=&h_{ij} \bar N_{R_i}\eta^\dagger\ell_{L_j}
+h_{ij}^\ast\bar\ell_{L_i}\eta N_{R_j}
+\frac{1}{2}\left(M_i\bar N_{R_i}N_{R_i}^c 
+M_i\bar N_{R_i}^cN_{R_i}\right), \nonumber \\
V&=&m_\phi^2\phi^\dagger\phi+m_\eta^2\eta^\dagger\eta
+\lambda_1(\phi^\dagger\phi)^2+\lambda_2(\eta^\dagger\eta)^2
+\lambda_3(\phi^\dagger\phi)(\eta^\dagger\eta) 
+\lambda_4(\eta^\dagger\phi)(\phi^\dagger\eta) \nonumber \\
&+&[\frac{\lambda_5}{2}(\phi^\dagger\eta)^2 +{\rm h.c.}],
\label{model}
\end{eqnarray}
where $\ell_{L_i}$ is a left-handed lepton doublet and $\phi$ is 
an ordinary Higgs
doublet. Interaction terms are written by using the basis under which
both matrices for Yukawa couplings of charged leptons and masses of 
the singlet neutrinos are real and diagonal.
Since the new doublet scalar $\eta$ is assumed to have no vacuum 
expectation value, the $Z_2$ symmetry forbids neutrino masses at tree level. 
We note that the lightest field with its odd parity is stable and then
its thermal relic behaves as DM in the universe. 

In this model we can define lepton number $L$ in two different ways, 
since $N_{R_i}$ and $\eta$ are introduced as the fields which couple
with $\ell_{L_i}$. In fact, if either $M_i\bar N_{R_i}N_{R_i}^c$ 
or $\lambda_5(\phi^\dagger\eta)^2$ decouples, the conserved lepton 
number can be defined for each case, respectively, such as
\begin{equation}
{\rm (i)}~ L(\eta)=0,\ L(N_{R_i})=1, \qquad
{\rm (ii)}~ L(\eta)=1,\ L(N_{R_i})=0.
\label{dl}
\end{equation}
The case (i) corresponds to the canonical seesaw model and the 
ordinary thermal leptogenesis is considered in this framework. 
The case (ii) is a new possibility allowed in this model.
If we take account of sphaleron interaction, both the baryon
number $B$ and the lepton number $L$ are violated but $B-L$ is
conserved as usual \cite{sph}. In each case, we find that 
$B-L$ is related to $B$ through the sphaleron interaction as
\begin{equation}
{\rm (i)}~ B=\frac{8}{23}(B-L), \qquad
{\rm (ii)}~ B=\frac{7}{19}(B-L).
\label{bl}
\end{equation} 
These are derived by using chemical equilibrium conditions \cite{chem}.

Now we consider the extension of the model at high energy regions
by introducing canonically normalized real singlet scalars $S_i$ 
to which the odd parity of $Z_2$ is assigned.
Their potential and interaction terms are assumed to be given by 
\begin{equation}
-{\cal L}_S= \sum_{i=1}^K\left(\frac{1}{2}m_i^2S_i^2
+\mu_iS_i\phi^\dagger\eta +{\rm h.c.}\right),
\label{ext}
\end{equation}
where $\mu_i$ is supposed to be complex.\footnote{The $Z_2$ symmetry can
not forbid quartic terms of $S_i$. However, we only assume that there
are no such terms here since the radiatively generated one can be
neglected for sufficiently small $\mu_i$.}
It is easy to find that the model defined by eq.~(\ref{model}) can be
obtained as the effective theory with 
$\lambda_5=\sum_i\frac{\mu_i^2}{m_i^2}$ after singlet scalars $S_i$ 
are integrated out.
Thus, if we adopt the definition (ii) for the lepton number 
in this extension, the origin of the lepton number violating 
$\lambda_5$ term in eq.~(\ref{model}) is found in ${\cal L}_S$. 
As long as this $\lambda_5$ term is out of thermal equilibrium at low energy 
regions, the $B-L$ is found to be a good symmetry there. 
We easily find that the lepton number violation due to $\lambda_5$ is crucial 
for the neutrino mass generation at weak scale.

Neutrino masses are generated through the one-loop effect by picking up
this lepton number violation associated with the electroweak symmetry
breaking. They can be expressed as
\begin{eqnarray}
{\cal M}^\nu_{ij}&=&\sum_{k=1}^3h_{ik}h_{jk}
\left[\frac{\lambda_5\langle\phi\rangle^2}
{8\pi^2M_k}\frac{M_k^2}{M_\eta^2-M_k^2}
\left(1+\frac{M_k^2}{M_\eta^2-M_k^2}\ln\frac{M_k^2}{M_\eta^2}\right)\right]
\nonumber \\
&\simeq&\left(\frac{\lambda_5}{10^{-10}}\right)\left(\frac{\bar h}{1.0}\right)^2
\left(\frac{1~{\rm TeV}}{M}\right)\times O\left(10^{-1}\right)~{\rm eV},
\label{nmass}
\end{eqnarray}
where $M_\eta^2=m_\phi^2+(\lambda_3+\lambda_4)\langle\phi\rangle^2$.
In the second line we put $h_{ik}\sim \bar{h}$ and $M_i, M_\eta\sim M$ for the
rough estimation.
If we remember that neutrino oscillation data requires that the neutrino
masses are $O(10^{-1})$~eV or less,\footnote{It should be 
noted that one of eigenvalues of this mass matrix is zero and also the
cosmological upper bound for the neutrino masses is 0.58~eV \cite{wmap}.} 
we find that $\lambda_5$ should be extremely small for the values of 
$\bar h$ and $M$ implicated in the second line of eq.~(\ref{nmass}).
The explanation of the smallness of $\lambda_5$ seems to be important 
for this neutrino mass generation mechanism to be natural. 
Several solutions for this question have been proposed in 
\cite{raddm2,susyrad2}. 
As we will see below, the present extension may give 
an alternative solution for this problem if $|\mu_i|\ll m_i$ is satisfied.
The effective coupling constant $\lambda_5$ is found to be closely related 
to the cosmological issues such as the inflation and the baryon number 
asymmetry in the universe. 
 
\section{Cosmological features of the model}
At first, we discuss the features of this extended model 
as an inflation model.  
If we assume that $S_i$ takes a large initial value such as 
$O(M_{\rm pl})$ where $M_{\rm pl}=(8\pi G)^{-1/2}$ is the reduced Planck mass,
the model behaves as the chaotic inflation model with multi-component 
inflaton. Various features for this kind of model have been discussed in
a lot of articles \cite{minf1}. 
We follow the results given in \cite{minf2}. 

The number of e-folds brought by $S_i$ is given as
\begin{equation}
N=M_{\rm pl}^{-2}\sum_{i=1}^K\int^{S_i^\ast}_{S_i^{\rm end}}
\left(\frac{V_i}{V_i^\prime}\right) dS_i\simeq\sum_{i=1}^K
\left(\frac{S_i^\ast}{M_{\rm pl}}\right)^2,
\end{equation}
where $V_i=\frac{1}{2}m_i^2S_i^2$.
$S_i^{\rm end}$ and $S_i^\ast$ stand for a field value at the end of
slow-roll inflation 
and a value at the time when cosmological scales $k=a_0H_0$
leave the horizon, respectively. 
Since $S_i^\ast\gg S_i^{\rm end}$ is satisfied, the number of e-folds $N$ is 
determined only by $S_i^\ast$ and it does not depend on the 
inflaton mass $m_i$. 
Using this e-folds $N$, the tilt of the scalar perturbation spectrum 
is given as
\begin{equation}
1-n_s=\frac{1}{N}+\frac{\sum_{i=1}^K m_i^4S_i^4}
{\left(\sum_{i=1}^Km_i^2S_i^2\right)^2}.
\end{equation}
If we note that $(\sum_i m_i^2S_i^2)^2>\sum_im_i^4S_i^4$ is satisfied
for $K\ge 2$, $n_s$ is found to become larger in the multi-component 
case compared with the single inflaton case. 
The ratio of tensor to scalar perturbation is given as
\begin{equation}
r=\frac{8}{N}.
\end{equation}
We note that $r$ is independent of the number of inflaton components 
and their masses. If all the masses of $S_i$ are equal, 
both $n_s$ and $r$ reduce to the ones of the single inflaton model 
\cite{minf2}. Both $n_s$ and $r$ is now constrained by the 7-year WMAP
data for the CMB observation as found in \cite{wmap}. 
Although the $m^2S^2$ type chaotic inflation with a single 
inflaton $S$ is marginally allowed for reasonable $N$ such as $N=50$ - 60, 
the situation becomes worse for the multi-component case 
since $n_s$ becomes larger compared to the
single inflaton case as discussed above 
(See Fig.~5 in the first paper of \cite{wmap}). 
Taking account of this, we confine our following study to the two 
component case with $m_1=m_2$ in eq.~(\ref{ext}).
  
The decay of inflaton $S_i$ occurs at $H\sim\Gamma_{S_i}<m_i$ 
through three scalars interaction $\mu_iS_i\phi^\dagger\eta$ 
in eq.~(\ref{ext}) where $H$ is the Hubble parameter and $\Gamma_{S_i}$
stands for the decay width of $S_i$.   
Since this decay width is given by 
$\Gamma_{S_i}=\frac{1}{4\pi}\frac{|\mu_i|^2}{m_i}$,
the decay $S_i\rightarrow \eta\phi^\dagger$ yields the thermal 
plasma with reheating temperature \cite{reh}
\begin{equation}
T_R\simeq 0.5g_\ast^{-1/2}\left(\frac{\mu_1}{m_1}\right)(m_1M_{\rm pl})^{1/2},
\label{tr}
\end{equation}
where we suppose $|\mu_1|\gg |\mu_2|$. In the present model 
we have $g_\ast=116$ as the relativistic degrees of freedom. 
If we note that this decay violates the lepton number defined as (ii) 
in eq.~(\ref{dl}), 
we find that the lepton number asymmetry can be induced in this process through 
the cross term between tree and one-loop diagrams as long as 
$\mu_i$ is complex.
The CP asymmetry induced in this process can be estimated as
\begin{eqnarray}
\varepsilon&\equiv&\sum_{i=1}^2\frac{\Gamma(S_i\rightarrow\eta\phi^\dagger)
-\Gamma(S_i\rightarrow\eta^\dagger\phi)}
{\Gamma(S_i\rightarrow\eta\phi^\dagger)
+\Gamma(S_i\rightarrow\eta^\dagger\phi)}
=\sum_{i,j}\frac{1}{8\pi}\frac{{\rm Im}(\mu_i^\ast\mu_j)^2}{m_i^2|\mu_i|^2}
\ln\frac{m_i^2+m_j^2}{m_j^2} \nonumber \\
&\simeq&\frac{|\lambda_5|\ln 2}{8\pi}\sin 2(\theta_1-\theta_2),
\label{cp}
\end{eqnarray}
where $\theta_i={\rm arg}(\mu_i)$ and we use 
$|\lambda_5|\simeq \frac{|\mu_1|^2}{m_1^2}$.

The lepton number asymmetry generated in the decay product 
$\eta$ is kept in thermal plasma if the effective $\lambda_5$ 
term in eq.~(\ref{model}) is out of thermal equilibrium 
until the sphaleron decoupling temperature. 
This condition is written as
\begin{equation}
|\lambda_5|<2\times 10^{-7}g_\ast^{1/4}\left(\frac{T}{1~{\rm TeV}}\right).
\end{equation}
Thus, when $|\lambda_5|$ is in this range, the lepton number
asymmetry can be estimated as the one induced through the decay of $S_1$ 
to $\eta\phi^\dagger$. It may be expressed as 
\begin{equation}
Y_L=\varepsilon Y_\eta(T_R)\simeq \varepsilon 
\frac{m_1S_1^{\rm end}}{s(T_R)}\simeq 10 \varepsilon\frac{T_R}{m_1}, 
\label{lasym}
\end{equation}
where $Y_L$ and $Y_\eta$ stand for the lepton number asymmetry and the
$\eta$ number in the co-moving volume. They are expressed 
as $Y_{L,\eta}=\frac{n_{L,\eta}}{s}$ by using the entropy density 
$s=\frac{2\pi^2}{45}g_\ast T^3$.
Since the washout of this generated lepton number asymmetry is caused by
the processes whose amplitudes are proportional to the tiny coupling
$\lambda_5$, it could be safely neglected. 

Combining eqs.~(\ref{bl}), (\ref{tr}), (\ref{cp}) and (\ref{lasym}),
we obtain the baryon number asymmetry $Y_B$ as
\begin{equation}
|Y_B|\simeq 5\times
 10^{-11}\left(\frac{\lambda_5}{10^{-10}}\right)^2
\left(\frac{10^5~{\rm GeV}}{T_R}\right)\sin 2(\theta_1-\theta_2).
\end{equation} 
Since the observed baryon number asymmetry in the universe is 
$Y_B=(0.7-0.9)\times 10^{-10}$ \cite{bbn}, 
it can be generated for the suitable values of $\lambda_5$ and $T_R$. 
For example, if we impose $\lambda_5=10^{-10}$ and $T_R=10^5$~GeV so as 
to generate suitable baryon number for the maximum 
value of $\sin 2(\theta_1-\theta_2)$, we have  
\begin{equation}
m_1=2\times 10^4~{\rm GeV}, \qquad |\mu_1|=0.2~{\rm GeV}.
\label{basym}
\end{equation} 
These results suggests that in this model the inflation should occur 
at a scale which is not far from the weak scale. 
In this extension the smallness of $\lambda_5$ which is crucial for the
small neutrino mass generation is explained as the 
nature of the inflaton sector.

The decay of $\eta$ also produces $N_{R_i}$ as the component of thermal plasma. 
Since the lightest singlet fermion $N_{R_1}$ is stable due to $Z_2$ symmetry, 
it behaves as DM. Thus, its relic abundance should satisfy 
$\Omega_{N_{R_1}}h^2=0.11$ which is obtained from observations of 
the WMAP \cite{wmap}. 
The relic abundance is fixed through the $N_{R_1}N_{R_1}$ 
annihilation caused by the $\eta$ exchange. 
This cross section can be expressed as \cite{raddm1,raddm3}
\begin{equation}
\langle\sigma v^2\rangle\simeq\frac{1}{12\pi}
\frac{M_1^2(M_1^4+M_\eta^4)}{(M_1^2+M_\eta^2)^4}
\sum_{i=e,\mu,\tau}|h_{i1}|^4\frac{6T}{M_1}.
\end{equation}
By using this cross section, the relic abundance can be 
approximately estimated as \cite{relic}
\begin{equation}
\Omega_{N_{R_1}}h^2=\frac{2.14\times 10^9z_f}{g_\ast^{1/2}m_{\rm pl}({\rm
 GeV})\langle\sigma v\rangle}, \quad
z_f=\ln\frac{0.038g m_{\rm pl}M_1\langle\sigma v\rangle}
{g_\ast^{1/2}z_f^{1/2}},
\end{equation}
where $z_f$ is defined as $z_f=M_1/T_f$ by using the freeze-out
temperature $T_f$ of $N_{R_1}$.
If this $\Omega_{N_{R_1}}h^2$ has the required value for the $\lambda_5$
estimated above, the model could give a consistent explanation for 
both the baryon number asymmetry and the DM abundance.  

In order to proceed this quantitative analysis, we need to introduce 
an additional assumption for the neutrino Yukawa couplings. 
Here we assume the following flavor structure for them \cite{raddm3}:
\begin{equation}
h_{ei}=0, \quad h_{\mu i}=h_{\tau i}=h_i~~(i=1,2); \qquad
h_{e3}=h_{\mu 3}=-h_{\tau 3}=h_3.
\label{fs}
\end{equation}
If we impose this structure, tri-bimaximal mixing is automatically
induced\footnote{Recent T2K and Double Chooz results suggest nonzero
$\sin\theta_{13}$ \cite{t13}. If we introduce a nonzero value for
$h_{ei}$ as a small perturbation to this flavor structure, nonzero 
$\sin\theta_{13}$ could be obtained \cite{radlept}. In the present study we
ignore this effect.} 
and then the neutrino oscillation data can be explained as long as
the following conditions are satisfied
\begin{equation}
|h_1|^2\Lambda_1+|h_2|^2\Lambda_2\simeq 
\frac{\sqrt{\Delta m_{\rm atm}^2}}{2}, \qquad
|h_3|^2\Lambda_3\simeq \frac{\sqrt{\Delta m_{\rm sol}^2}}{3},
\label{c-oscil}
\end{equation}
where $\Delta m^2_{\rm atm}$ and $\Delta m^2_{\rm sol}$ 
represent the squared neutrino mass difference required by the
atmospheric neutrinos and solar neutrinos \cite{oscil}.  
$\Lambda_i$ is defined as
\begin{equation}
\Lambda_i=\frac{\lambda_5\langle\phi\rangle^2}
{8\pi^2M_k}\frac{M_k^2}{M_\eta^2-M_k^2}\left(1+\frac{M_k^2}
{M_\eta^2-M_k^2}\ln\frac{M_k^2}{M_\eta^2}\right).
\end{equation} 
If we use $\lambda_5$ derived from $m_1$ and $\mu_1$ given 
in eq.~(\ref{basym}) and impose the neutrino 
mass constraints (\ref{c-oscil}), we can estimate the DM abundance.
Here we proceed this analysis only for typical parameters such as
\begin{equation}
|\lambda_5|=2.4\times 10^{-10}, \quad M_2=4~{\rm TeV}, \quad  M_3=6~{\rm TeV}.
\end{equation}    

\input epsf
\begin{figure}[t]
\begin{center}
\epsfxsize=7.5cm
\leavevmode
\epsfbox{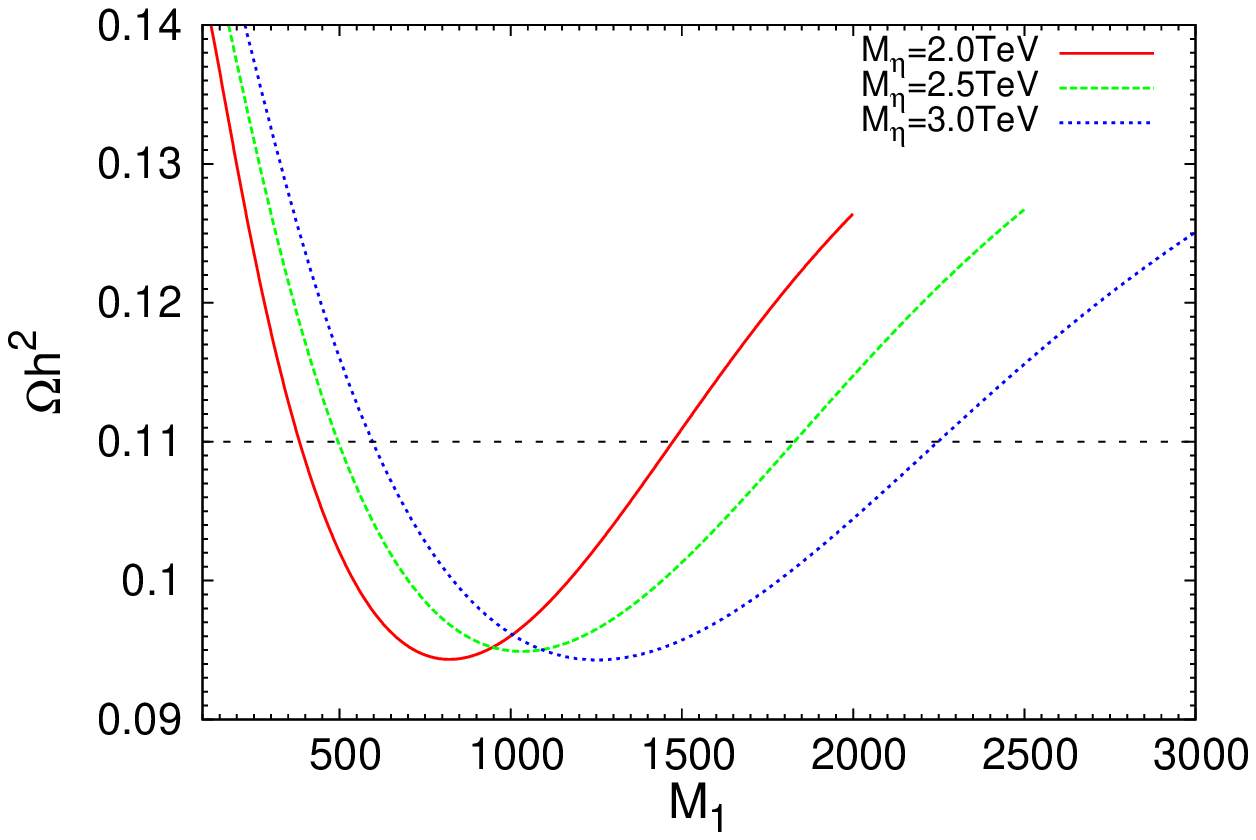}
\hspace*{5mm}
\epsfxsize=7.5cm
\leavevmode
\epsfbox{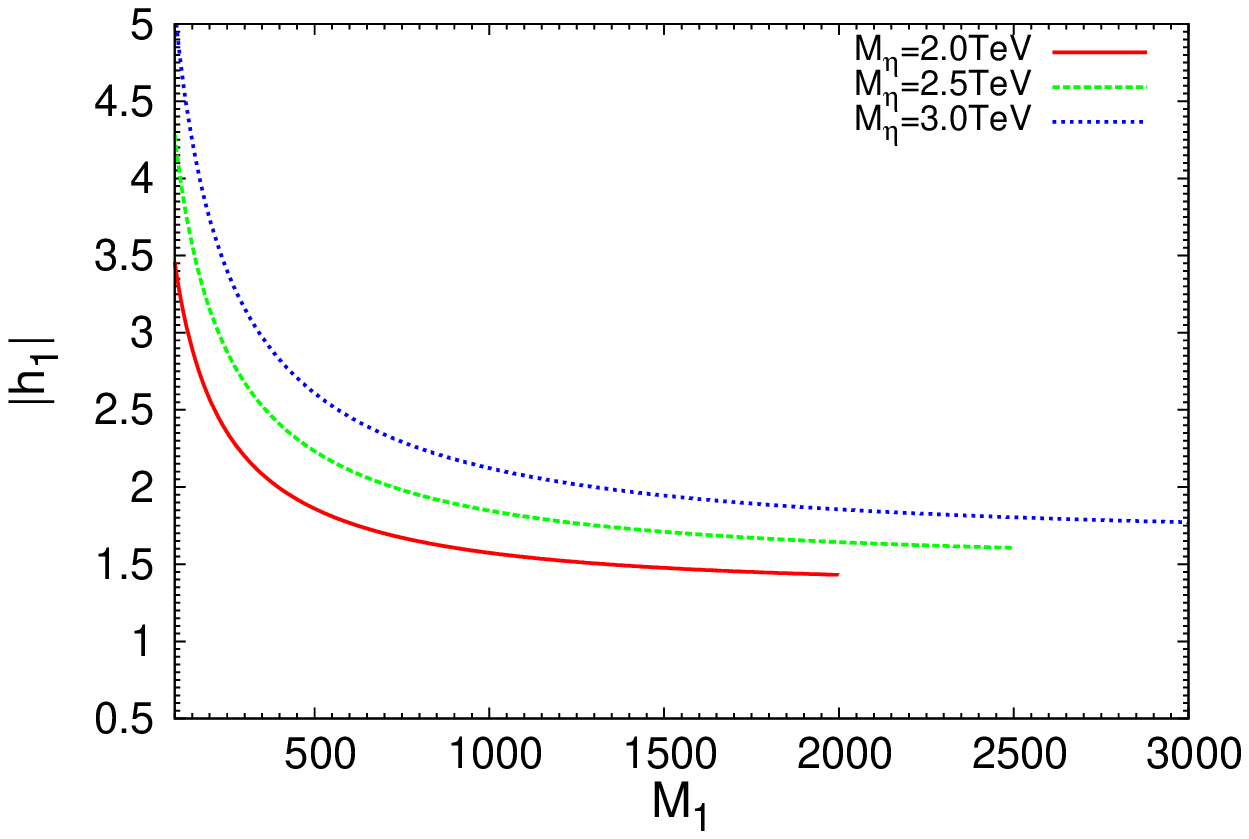}
\end{center}
\vspace*{-3mm}

{\footnotesize {\bf Fig.~1}~~The left frame shows the $N_{R_1}$ relic abundance 
as a function of $M_1$. The right frame shows the neutrino Yukawa 
coupling $|h_1|$ as a function of $M_1$.
Each line is plotted for $M_\eta=2$, 2.5 and 3~TeV.}  
\end{figure}

In the left frame of Fig.~1, we plot the $N_{R_1}$ relic abundance
$\Omega_{N_{R_1}}h^2$ as a function of $M_1$ for some values of $M_\eta$.
In this calculation we assume $h_2=h_1$, for simplicity.
If a value of $\lambda_5$ becomes smaller, the neutrino Yukawa coupling 
$|h_1|$ becomes larger as expected from eq.~(\ref{nmass}) and then 
$\Omega_{N_{R_1}}h^2$ takes a smaller value. 
We can check that $\lambda_5$ should satisfy 
$|\lambda_5|~{^<_\sim}~2.6\times 10^{-10}$
to reduce $\Omega_{N_{R_1}}h^2$ to the observed value for the present
parameter setting.
As long as this bound is satisfied, the required $N_{R_1}$ relic 
abundance can be obtained consistently with the neutrino mass constraints.
In right frame of Fig.~1, we plot the neutrino Yukawa coupling 
$|h_1|$ as a function of $M_1$ by applying the condition in eq.~(\ref{c-oscil}) 
for the same parameters used in the left frame of Fig.~1. 

\begin{figure}[t]
\begin{center}
\begin{tabular}{|c|c|c|c|c|}\hline
$M_\eta$~(TeV) & $M_1$~(TeV) & $|h_1|$ & $Br(\mu\rightarrow e\gamma)$ &
 $Br(\tau\rightarrow \mu\gamma)$ \\ \hline
2  &0.38  &2.02  & $0.36\times 10^{-12}$  & $9.5\times 10^{-8}$  \\ \hline
2  &1.55  &1.47 & $1.5\times 10^{-12}$  & $0.34\times 10^{-8}$   \\ \hline
2.5&0.49  &2.25  &$0.42\times 10^{-12}$   &$10\times 10^{-8}$   \\ \hline
2.5&1.9  &1.65  & $0.42\times 10^{-12}$  & $0.46\times 10^{-8}$   \\ \hline
3  &0.59  &2.47  & $0.46\times 10^{-12}$  & $11\times 10^{-8}$   \\ \hline
3  &2.3  &1.82  & $0.46\times 10^{-12}$  & $0.55\times 10^{-8}$   \\ \hline
\end{tabular}
\end{center}
\vspace*{2mm}
{\footnotesize{\bf Table 1}~~Predicted values of the neutrino Yukawa 
coupling $|h_1|$ and the branching ratio of the lepton flavor 
violating processes by imposing the neutrino oscillation data and the DM
 relic abundance on the model.}
\end{figure} 

Combining these results, we can find the values of Yukawa couplings 
$h_1$ and $h_3$ which give the required value of $\Omega_{N_{R_1}}h^2$.  
These values are shown in Table~1.
Since the neutrino Yukawa couplings are large, the model could be 
severely constrained by the lepton flavor violation processes such as
$\tau\rightarrow\mu\gamma$ and $\mu\rightarrow e\gamma$ \cite{raddm1}. 
The branching ratio of these processes are given by
\begin{eqnarray}
&&{\rm Br}(\tau\rightarrow\mu\gamma)\simeq 
\frac{0.51\alpha}{64\pi(G_FM_\eta^2)^2}
\left[|h_1|^2\left\{F_2\left(\frac{M_1^2}{M_\eta^2}\right)+
F_2\left(\frac{M_2^2}{M_\eta^2}\right)\right\}
-|h_3|^2F_2\left(\frac{M_3^2}{M_\eta^2}\right)\right]^2, \nonumber\\
&&{\rm Br}(\mu\rightarrow e\gamma)\simeq \frac{3\alpha}{64\pi(G_FM_\eta^2)^2}
\left[|h_3|^2F_2\left(\frac{M_3^2}{M_\eta^2}\right)\right]^2, 
\label{flfv}
\end{eqnarray}
where $F_2(r)$ is defined as
\begin{equation}
F_2(r)=\frac{1-6r+3r^2+2r^3-6r^2\ln r}{6(1-r)^4}.
\label{f2}
\end{equation}
The predicted values of these branching ratio for the obtained parameters
are also given in Table~1. The present upper bounds for 
$Br(\mu\rightarrow e\gamma)$ and $Br(\tau\rightarrow \mu\gamma)$ 
are $2.4\times 10^{-12}$ and $4.4\times 10^{-8}$, respectively
\cite{lfv}. One may consider that the former gives much severe 
constraint on the model. 
However, the situation is different here.
Although the neutrino Yukawa coupling $|h_1|$ have to take large values 
to reduce the $N_{R_1}$ relic abundance, 
the assumed flavor structure (\ref{fs}) can
successfully suppresses $\mu\rightarrow e\gamma$ \cite{raddm2}.
As found from eq.~(\ref{flfv}), this process depends 
only on $|h_3|$ which can be much smaller than $|h_1|$.
On the other hand, since $\tau\rightarrow \mu\gamma$ depends on $|h_1|$,
it could impose severer constraint on this model. 
Anyway, we can find consistent parameters for which these bounds are 
satisfied successfully as shown in Table 1.

Finally, we should remark on other phenomenological problems which could appear
in relation to the present neutrino mass generation.
First, we note that there is one-loop contribution to the muon $g-2$
through the interactions given in eq.~(\ref{model}) \cite{raddm3}.
By using eq.~(\ref{f2}), it can be written as \cite{bmeg}
\begin{eqnarray}
\delta a_\mu&=&\sum_{k=1}^3\frac{|h_{\mu k}|^2}{(4\pi)^2}
\frac{m_\mu^2}{M_\eta^2}F_2\left(\frac{M_k^2}{M_\eta^2}\right)
\simeq 7.1\times 10^{-11} \left(\frac{2~{\rm TeV}}{M_\eta}\right)^2
\left(\frac{|h_1|}{2}\right)^2 {\cal O}, \nonumber \\
{\cal O}&=&F_2\left(\frac{M_1^2}{M_\eta^2}\right)
+F_2\left(\frac{M_2^2}{M_\eta^2}\right)
+\frac{|h_3|^2}{|h_1|^2}F_2\left(\frac{M_3^2}{M_\eta^2}\right),
\end{eqnarray} 
where we use eq.~(\ref{fs}) in this derivation. 
Since ${\cal O}$ is less than 0.5 here, the predicted value of 
$\delta a_\mu$ is found to be two orders of 
magnitude smaller than the present experimental value 
$\delta a_\mu=(30.2\pm 8.7)\times 10^{-10}$ \cite{mg2}. 
We need other contributions to saturate this discrepancy 
if we take it seriously.

Second, the model could have a problem relevant to the CP phases,
which eventually appears when we consider the leptogenesis. 
As in case of the muon $g-2$, the interactions introduced to 
generate the neutrino masses could also contribute to the electric 
dipole moment of an electron (EDME) through loop diagrams 
if they violate the CP invariance.
As found from the neutrino mass formula (\ref{nmass}), the phase of
$\mu_1$ contributes to the MNS matrix $U$ as the overall Majorana phase.
We note that in the EDME loop diagrams the MNS matrix elements appear 
at lepton verteces with a $W^\pm$ line and also an $\eta_0$ line in pairs as 
$U_{ij}U_{i^\prime j^\prime}^\ast$. This means that the CP phase of $\mu_1$
which is relevant to the leptogenesis does not contribute to the EDME.
Although the EDME could be induced by the diagrams with more than
two loops as a result of the CP phases of neutrino Yukawa couplings,
it could satisfy the present experimental upper bound \cite{edm} easily
by fixing the parameters irrelevant to the leptogenesis suitably.  
The constraint from the EDME does not contradict with the present 
leptogenesis scenario.
 
\section{Summary}
We have considered an extension of 
the radiative neutrino mass model as an inflation model.
In this extension the original model is obtained as
the effective theory by integrating out the inflaton field.
The lepton number violating term which plays a crucial role
in the radiative neutrino mass generation appears in relation to 
the inflaton sector. As a result, the neutrino masses are
closely related to the reheating temperature and also the baryon 
number asymmetry in the universe through the parameter $\lambda_5$. 
The model might also give a unified picture for the explanation 
of the origin of DM other than these.    
Since the resulting reheating temperature is low enough
to escape the gravitino problem \cite{gravitino}, 
the supersymmetric extension of the present model along the framework 
given in \cite{susyrad1,susyrad2} could be an interesting subject.
It will be discussed elsewhere.

\section*{Acknowledgement}
This work is partially supported by a Grant-in-Aid for Scientific
Research (C) from Japan Society for Promotion of Science (No.21540262)
and also a Grant-in-Aid for Scientific Research on Priority Areas 
from The Ministry of Education, Culture, Sports, Science and Technology 
(No.22011003).

\newpage
\bibliographystyle{unsrt}

\end{document}